# Quantifying human mixing patterns in Chinese provinces outside Hubei after the 2020 lockdown was lifted

Yining Zhao[1], Samantha O'Dell[2], Xiaohan Yang[3], Jingyi Liao[4], Kexin Yang[1], Laura Fumanelli[2], Tao Zhou[5], Jiancheng Lv[1], Marco Ajelli[2,#], Quan-Hui Liu[1,#,*]

[1] College of Computer Science, Sichuan University, Chengdu, China
[2] Laboratory for Computational Epidemiology and Public Health, Department of Epidemiology and Biostatistics, Indiana University School of Public Health, Bloomington, IN, USA
[3] Institute for Applied Computational Science, Harvard University, Cambridge, MA, USA
[4] Shenzhen International Graduate School, Tsinghua University, Shenzhen, China
[5] Big Data Research Center, University of Electronic Science and Technology of China, Chengdu, China,

[#] Senior author.

[*] Corresponding author: quanhuiliu@scu.edu.cn

**Abstract**

Contact patterns play a key role in the spread of respiratory infectious diseases in human populations. During the COVID-19 pandemic the regular contact patterns of the population has been disrupted due to social distancing both imposed by the authorities and individual choices. Here we present the results of a contact survey conducted in Chinese provinces outside Hubei in March 2020, right after lockdowns were lifted. We then leveraged the estimated mixing patterns to calibrate a model of SARS-CoV-2 transmission, which was used to estimate different metrics of COVID-19 burden by age. Study participants reported 2.3 contacts per day (IQR: 1.0-3.0) and the mean per-contact duration was 7.0 hours (IQR: 1.0-10.0). No significant differences were observed between provinces, the number of recorded contacts did not show a clear-cut trend by age, and most of the recorded contacts occurred with family members (about 78%). Our findings suggest that, despite the lockdown was no longer in place at the time of the survey, people were still heavily limiting their contacts as compared to the pre-pandemic situation. Moreover, the obtained modeling results highlight the importance of considering age-contact patterns to estimate COVID-19 burden.

## 1. Introduction

The COVID-19 pandemic has caused countries around the globe to adopt unprecedented measures to combat the spread of the disease. China has adopted lockdown and social distancing strategies to decrease the newly diagnosed infection rate of SARS-CoV-2 [1]. As a result, provinces outside Hubei were able to quickly contain the spread of infection [2]. Further, a spontaneous behavioral change regarding human mixing patterns has been observed throughout populations [3-9]. After the lockdowns were lifted, many individuals were still hesitant to resume normal activities. This behavioral change is a significant indicator of how mixing patterns have changed throughout the population during the pandemic.

Before the COVID-19 pandemic, many studies have focused on age-mixing patterns to understand the spread of infectious diseases [10-14]. Those patterns, however, are representative of a pre-pandemic situation and have little in common with life during this pandemic [15]. As shown by human mobility data (e.g., Google, Apple, Yandex, Weibo [3,16-18]), activity levels have dramatically changed over the course of the pandemic, compared to pre-pandemic conditions. However, this data does not provide any estimate of the mixing patterns by age contacts or other descriptive measures needed to fully understand COVID-19 epidemiology. As a result, only a handful of studies have taken on this approach to study how the age and duration of mixing patterns have changed during the pandemic as compared to pre-pandemic conditions [6-8].

The aim of this study is to estimate age-mixing patterns relevant for the spread of SARS-CoV-2 in Chinese provinces outside Hubei over the period of March 3-23, 2020 through survey-based contact diaries. During this period, lockdowns had just lifted, but social distancing was still in place. Individuals who participated in the survey were asked to record all persons they were in contact with over a 24-hour period, with contact being defined as exchanging more than three words in physical presence or having a physical contact. Participants were then asked to define each contact by age, sex, employment status, relationship to the contact, the social setting in which the contact took place, and the duration of the interaction. We then leveraged a mathematical model calibrated using the estimated mixing patterns to provide an insight on their role in SARS-Cov-2 transmission as opposed to a homogeneous mixing of the population.

## 2. Methods

### 2.1. Survey on contact patterns

To estimate age-mixing patterns, we performed a contact survey by Wenjuanxing, a professional online survey platform widely adopted in China [19]. Participants were recruited to the survey between March 3, 2020 and March 23, 2020, as this was the time that lockdowns were lifted in most provinces, although other non-pharmacological interventions (NPIs) were still in place. Those interventions included school closure, promotion of remote working, temperature taking to access public places, wearing masks in public, prohibiting mass gatherings, and scanning HealthQR codes. Hubei Province was not included in this analysis, as Wuhan (its largest city) remained in lockdown during this time.

When the survey was first given to the participants, the background and purpose of the survey were explained in detail. Confidentiality was also guaranteed to each participant for all personal information collected. Additionally, participants signed a confirmation of authenticity to ensure that all information they provided was correct. They also agreed to have their IP addresses automatically taken from the platform to record the Chinese Province where they were based at the time of the survey.

The survey consisted of two areas of questioning: 1) demographic information; and 2) social contact information. The demographic information questions regarded participants' age, sex, province of residence, as well as the age and sex of all members of the participants' households.

The contact diary was then assigned to the participants to record all individuals they came into contact with in a 24-hour time period. A contact was defined as either: 1) a two-way conversation that involved at least 3 words in the physical presence of another person (conventional contact), or 2) a direct physical contact (e.g., a handshake, hug, kiss) [10]. As a result, only individuals that the participants interacted with that met these conditions were included in the contact diaries. For each contact, the participants recorded the age (or age range if age was unknown); the relationship of the contact (household member, other relative, classmate/colleague, other schoolmate, other); where the contact took place (house, work, school, leisure, transport, other); and how many minutes they were in the same room or environment of the contact.

**2.2. Ethics statement**

The study was approved by the Clinical Trials and Biomedical Ethics Committee of the West China Hospital, Sichuan University (No. 2020190). Data were anonymized, and informed consent was waived.

**2.3. Statistical analysis**

To quantify the average number of contacts and contact duration for each age group and to construct contact matrices by age for the actual population of China [20], bootstrap sampling [21] with replacement was utilized.

We estimated the mean number of contacts and their mean duration from the survey, both irrespective of age and for each age group. We then scaled to the actual population of China: the mean number of contacts weighted by the age distribution of the population gives the population-adjusted mean number of contacts. By an analogous procedure we obtained the population-adjusted mean duration of contacts. Mann Whitney with the corresponding independent sample t-test was used to determine whether differences (e.g., by province) in the number of contacts and duration were statistically significant. Bootstrap sampling with replacement was utilized to estimate statistics (such as standard error) on contact patterns.

**2.4. Modeling SARS-CoV-2 transmission**

To investigate the impact of contact patterns on SARS-CoV-2 spread as well as COVID-19 burden, we developed an age-structured compartmental SIR model [22]. The population is divided into three compartments: susceptible (S), representing individuals who can acquire the infection; infectious (I), representing individuals who are infected and able to transmit the infection; and removed (R), representing individuals who are immune to the infection. Each compartment is divided into 18 5-year age groups (0–4, 5–9, … , 80-84,85+). Susceptible individuals are exposed to an age-specific force of infection that depends on the number of infectious individuals of a given age, the matrix of contacts regulating the intensity

(i.e., number and duration) of contacts with individuals of a given age per unit of time, and the transmission rate per contact. Infectious individuals move to the removed compartment according to a recovery rate. This process is regulated by the following system of ordinary differential equations:

$$\dot{S}_i = -S_i \beta \sum_{j=1}^{n} M_{g(i)g(j)} T_{g(i)g(j)} \frac{\sum_{k=1,\ldots,m} I_k \delta(g(k),g(j))}{\sum_{k=1,\ldots,m} N_k \delta(g(k),g(j))}$$

$$\dot{I}_i = S_i \beta \sum_{j=1}^{n} M_{g(i)g(j)} T_{g(i)g(j)} \frac{\sum_{k=1,\ldots,m} I_k \delta(g(k),g(j))}{\sum_{k=1,\ldots,m} N_k \delta(g(k),g(j))} - \gamma I_i$$

$$\dot{R}_i = \gamma I_i$$

where,

- $S_i$ represents the number of susceptible individuals in age group i;
- $I_i$ represents the number of infectious individuals in age group i;
- $R_i$ represents the number of removed individuals in age group i;
- $N_i$ represents the total number of individuals in age group i (i.e., $N_i = S_i + I_i + R_i$);
- $g( )$ is a function used to map the n=18 age groups used for the population to the *m*=6 age groups used in the contact matrix. Therein,

$$g(k) = \begin{cases} 1, & if\ k \leq 4, \\ 2, & if\ k = 5, \\ 3, & if\ k = 6, \\ 4, & if\ k = 7, \\ 5, & if\ k = 8,9, \\ 6, & if\ k > 9; \end{cases}$$

- $M_{g(i)g(j)}$ is the mean number of contacts that an individual in age group g(i) has with individuals in age group g(j);
- $T_{g(i)g(j)}$ is the mean duration of a contacts that an individual in age group g(i) has with individuals in age group g(j);
- $\delta(g(k), g(j))$ is the Dirac delta function (i.e., it is equal to 1 if g(k)=g(j); 0 otherwise);
- $\beta$ is the per-contact transmission rate per day;
- $\gamma$ is the recovery rate, which corresponds to the inverse of the generation time in an SIR model [23,24], and it is set to 5.1 days [2].

A key parameter regulating the spread of the infection is the basic reproduction number, $R_0$, representing the average number of infections generated by a typical primary infector in a fully susceptible population over the whole duration of their infectious period [22]. The basic reproduction number is computed by the next-generation approach [25].

Along with the infection transmission model, we developed a disease burden model. This model takes as input the number of infected individuals by age estimated by the transmission model and applies age-specific risks to estimate the distributions by age of some major indicators of disease burden: symptomatic infections, hospital admissions, and deaths. The values of these age-specific risks are taken from the literature [26, 27] and reported in Tab. S1. We then run the model using three assumptions on the mixing patterns of the population: homogeneous mixing, contact matrix (based on the recorded number of contacts by age only), and contact and duration matrix (based on the recorded number of contacts by age and their duration).

## 3. Results

**Description of the sample.** In total, we collected 748 diaries; 394 participants were excluded from the study. Of these excluded participants, 311 did not have complete contact or residential information, 14 did not reside in Mainland China outside Hubei, and 1 responded to the survey outside of the study period. In the remaining 422 clean diaries, 200 diaries are from Sichuan. Outside Sichuan, only participants from the provinces with at least 10 diaries were kept, which brings the total number diaries analyzed in this study to 354 (Tab. S2). No significant difference in both contact numbers and contact duration with Sichuan were found for these provinces (Mann Whitney test, all p-values > 0.01, except for one outlier). Shanxi Province exhibited an average contact duration of 9.7 hours as compared to an overall average of 7.0 hours.

**Number of contacts and duration.** A total of 828 contacts were analyzed in this study. The mean number of daily contacts was 2.3 (interquartile range, IQR: 1.0-3.0) (Tab. 1). These estimates are similar to estimates taken in February 2020 for Wuhan, which reported 2.0 contacts on average, and Shanghai, resulting in 2.3 average contacts [7]. The overall mean contact duration was 7.0 hours (IQR: 1.0-10.0). These estimates are also shorter than the average contact person-hours that were reported in Hong Kong in a 2015/16 online questionnaire, which resulted in 9.3 total contact person-hours [12]. The shorter contact duration found here is likely due to the specific time periods this survey was conducted, suggesting that at the time of the survey, despite the lockdown being lifted, people were still spending less time contacting other people.

**Table 1** Number and duration of recorded contacts per participant.

| Characteristics | Number of Participants | Contact number Mean (IQR) | Contact duration Mean (IQR) |
|---|---|---|---|
| **Overall** | 354 | 2.3 (1.0, 3.0) | 7.0 (1.0, 10.0) |
| **Age of participant** | | | |
| 0-19 | 65 | 2.3 (2.0, 3.0) | 7.5 (1.2, 12.0) |
| 20-24 | 189 | 2.4 (1.0, 3.0) | 7.0 (1.8, 10.0) |
| 25-29 | 93 | 2.0 (1.0, 3.0) | 7.9 (0.8, 10.0) |
| 30-34 | 39 | 2.8 (1.0, 4.0) | 5.9 (0.4, 11.0) |
| 35-39 | 20 | 3.1 (2.0, 4.0) | 6.6 (2.2, 10.0) |
| 40+ | 17 | 2.0 (1.0, 3.0) | 5.4 (0.3, 7.8) |
| **Sex of participant[1]** | | | |
| Male | 189 | 2.4 (1.0, 3.0) | 6.4 (1.0, 10.0) |
| Female | 164 | 2.3 (1.0, 3.0) | 6.9 (1.4, 10.0) |
| **Household size** | | | |
| 1 | 20 | 1.7 (0.0, 2.0) | 2.9 (0.0, 6.0) |
| 2 | 108 | 1.6 (1.0, 2.0) | 6.1 (0.5, 10.0) |
| 3 | 121 | 2.2 (2.0, 2.0) | 6.9 (1.8, 10.0) |
| 4 | 76 | 2.9 (2.0, 3.0) | 7.4 (2.0, 10.6) |
| 5 | 19 | 4.1 (4.0, 4.0) | 7.5 (3.0, 9.6) |
| 6+ | 10 | 2.3 (1.0, 3.0) | 7.0 (1.0, 10.0) |
| **Provinces** | | | |
| Sichuan | 200 | 2.3 (1.0, 3.0) | 6.5 (0.8, 10.0) |
| Chongqing | 20 | 2.8 (1.0, 4.0) | 7.6 (2.5, 12.0) |
| Shandong | 19 | 2.1 (1.0, 2.0) | 7.0 (0.5, 11.7) |
| Hebei | 17 | 2.1 (2.0, 2.0) | 7.4 (2.0, 10.0) |
| Henan | 16 | 2.7 (2.0, 3.0) | 5.1 (0.8, 11.0) |
| Zhejiang | 14 | 2.4 (2.0, 3.0) | 5.8 (0.9, 12.0) |
| Yunnan | 13 | 3.1 (1.0, 3.0) | 5.5 (1.5, 9.6) |
| Fujian | 13 | 2.2 (2.0, 2.0) | 6.3 (1.0, 8.2) |
| Hunan | 12 | 1.9 (1.0, 3.0) | 7.0 (1.0, 11.3) |
| Guangdong | 10 | 2.2 (1.0, 3.0) | 7.2 (2.0,10.0) |
| Jiangxi | 10 | 3.0 (2.0, 4.0) | 6.6 (2.2,10.0) |
| Shanxi | 10 | 2.8 (2.0, 3.0) | 9.7 (5.0,12.5) |

[1] Note that one participant refused to fill in their gender.

**Contact patterns by gender, age, relationship, and location.** Individuals with different genders demonstrated no apparent distinctions in their contact numbers and duration, resulting in 2.3-2.4 contacts and 6.4-6.9 hours respectively (Mann Whitney test, p-values>0.01). The average number of contacts increased as the household size became larger, rising from 1.7 with a household size of one to 6.6 for a household size of six or more. The average contact duration also exhibited the same pattern, increasing from 2.9 for a household of one to 9.0 for households with six or more members, comparable with reference [12]. The mean number of contacts by age shows only slight differences in individuals aged from 30-34 and 35-39. Individuals in the 30-34 age group and individuals in the 35-39 age group exhibited daily mean contacts as 2.8 and 3.1 respectively, which is slightly higher than the daily mean contacts for individuals in the other age groups (Table 1). Since the numbers of participants in the six age groups were unbalanced, we performed bootstrap sampling with replacement of survey participants weighted by the age distribution of the actual population of China from a 2010 census [21]. We then recalculated the population-adjusted mean number of contacts in this fashion. It can be discerned that there were no obvious differences obtained in the distribution of contact numbers after bootstrap sampling (Fig. 1A).

In Figure 1B, individuals in age groups 0-19 and 20-24 have the lowest numbers of same-age contacts, while individuals in the 40+ age group had the highest number. This finding was possibly explained by the fact that these age groups are students in the population and their school semester was postponed during the time of the study. On the other hand, a considerable part of the people in the 40+ age group are workers in the population and some of those businesses opened again after some of the initial interventions were lifted. The obtained result is also due to the different size by age group, with individuals aged 40+ years far outweighing any other age group (42.6% of the total Chinese population).

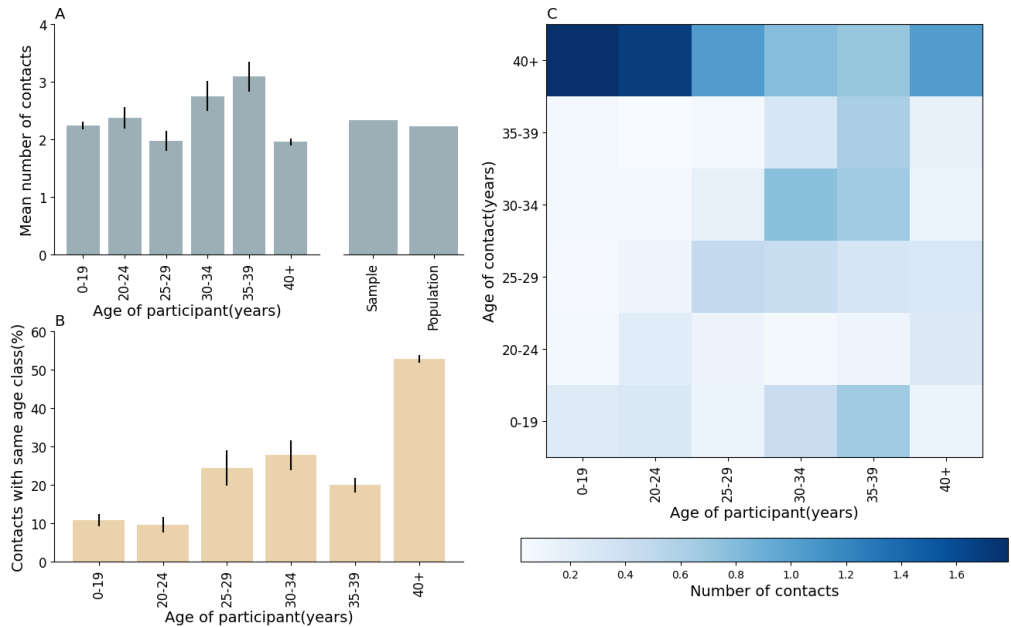

**Figure 1**: **Daily number of contacts and contacts by age. A**. Mean number of contacts by age group of study participants. The standard error in each age group is obtained from bootstrap sampling according to the age distribution of Chinese population from the 2010 census. The two bars on the right show the mean number of contacts irrespective of the age of study participant in the sample and by adjusting for the age structure of the actual Chinese population. **B**. Fraction of contacts with same age class as participants. **C**. Overall contact matrix by age group.

Mixing patterns by age can be summarized in a contact matrix, whose elements represent the mean number of contacts that an individual in a given age group has with individuals in other age groups (Fig. 1C). The resulting contact matrix by age is characterized by the presence of a main diagonal representing contacts with individuals in the same age group. Contacts between school age individuals are lower than 1 per day on average; in fact, most schools were still closed in the analyzed time frame.

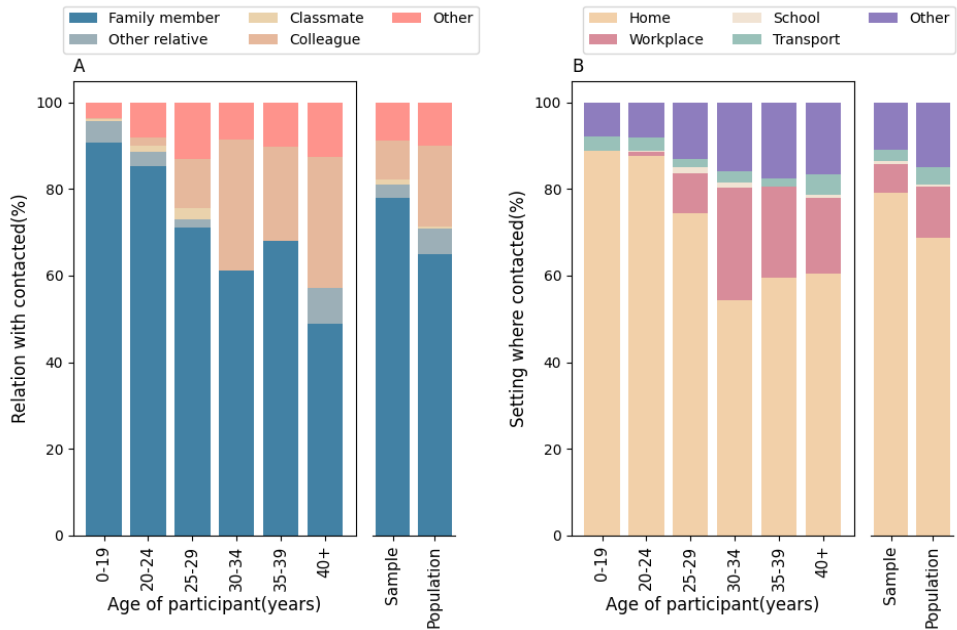

**Figure 2**: **Relationship between contacts and location of contacts. A**. Probability distribution of the relationship between the study participant of a given age group and the contacted individual. The bars on the right show the probability distribution of the relation between study participant irrespective of age and the contacted individual in the sample and by adjusting for the age structure of the Chinese population. **B**. As A, but showing the location where contacts took place.

As for the relationship between contacts, 77.8% of recorded contacts occurred between family members, 3.2% with other relatives, 1.2% classmates, 8.9% with colleagues, and 8.8% with other individuals (Fig. 2A). This figure highly differs from what is generally estimated in pre-pandemic studies [10,13,14,28,29], but consistent with what was observed during the early phase of the COVID-19 pandemic [7,8,15]. Despite the lockdown being lifted at the time of our survey, our results show that contacts were predominantly occurring within households. This trend is more marked for younger individuals: for individuals aged 0-19 years, 90.7% of contacts occurred with household members, while this figure decreases to 44.8% for individuals aged 40 years or more. This is likely associated with a gradual resumption of in-person work; in fact, depending on the age group, the proportion of contacts with work colleagues varies in the range 11.2%-34.5% (Fig. 2A). The fraction of contacts with other relatives varied between 5% and 12.1%, with the larger fraction found in the age groups 0-19 and 40+.

We estimated that the household was the most common place where contacts took place (79.1%), followed by "other places" (10.8%), workplace (6.8%), transport (2.8%), and school (0.6%) (Fig. 2B). Individuals aged 40+ years made about 20% of their contacts in the workplace.

**Contact duration.** In this survey, the contact duration per contact was also collected. As shown in Fig. 3A, study participants presented an average contact duration per contact from 5.4 to 7.5 hours. Individuals aged 40+ years showed the shortest average duration per contact, while participants in age group 25-29 show the longest average duration per contact. Similarly, no significant differences can be seen in the duration per contact after adjusting for the actual age distribution of the Chinese population (Fig. 3A). Participants over 40 had spent the most amount of time with same-age contacts, while individuals in age groups 0-19 and 20-24 spent the lowest amount of time with same-age contacts (Fig. 3B). The average durations per contact between individuals of different age groups are reported in Fig. 3C. As for the relationship between contacts, the duration that participants had contact with family members comprised the majority of the total time, mainly at 84.8% (Fig. 3D), followed by time spent with colleagues for 2.8 hours. Both young and old people tended to spend more time with other relatives.

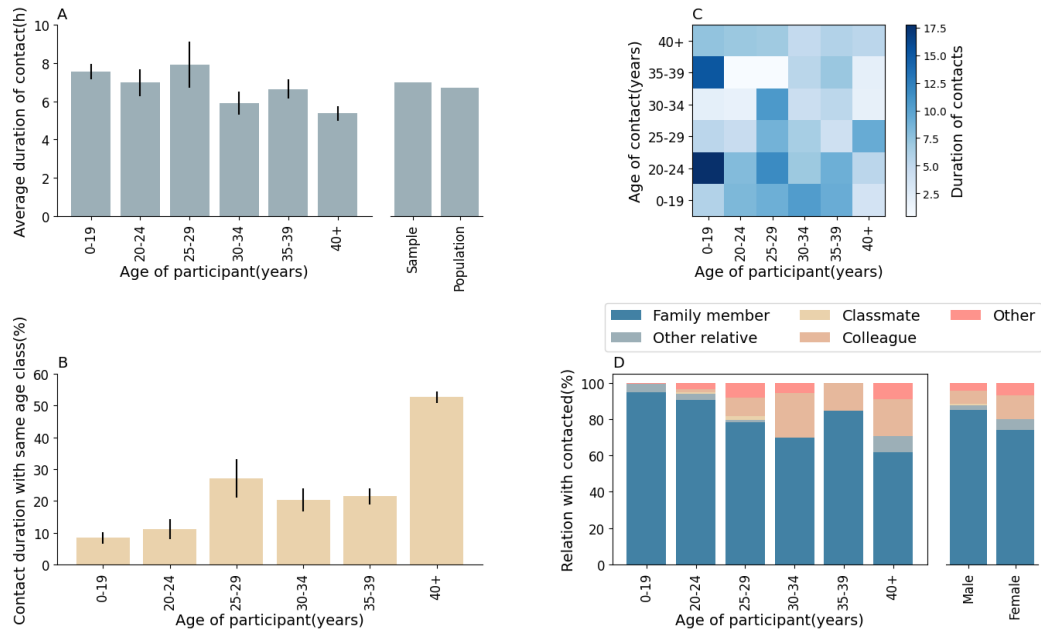

**Figure 3**: **Daily average duration of contacts by age. A**. Daily average duration of contacts by age group of study participants. The standard error in each age group is obtained from bootstrap sampling according to the age distribution of the Chinese population from the 2010 census. The bars on the right show the daily average duration per contact between study participant irrespective of age and the contacted individual in the sample and by adjusting for the age structure of the Chinese population. **B**. Fraction of contact duration with same age class individuals. **C**. Overall contact duration matrix by age groups. **D**. Probability distribution of the relation between study participant of a given age group and contacted individuals.

**Impact of contact patterns on COVID-19 burden.** We investigated the impact of contact heterogeneity on the spread of COVID-19 by comparing three different models. Model 1 assumed homogeneous mixing of the population; Model 2 assumed contacts were heterogeneous by age according to the survey (i.e., the estimated contact matrix) but the duration was homogeneous (i.e., $T_{ij}=1$, where i, j=1, 2, …, 6); Model 3 assumed that contacts and their duration were heterogeneous by age according to the survey (i.e., contact matrix and contact duration matrix were used).

We simulated SARS-CoV-2 spread assuming a reproduction number of 1.3 and we let the three models run until 1,000 cumulative infections were reached (in a population of 20 million individuals); we then analyzed the distributions by age of the major indicators of disease burden. The reproduction number and the number of infections were selected to describe a situation that resemble what was observed in Chinese provinces outside Hubei in the period right antecedent to our contact survey [2].

The simulation results show, for each indicator of the disease burden, marked differences between the homogeneous mixing model (Model 1) and the two models considering heterogeneous contacts by age (Model 2 and 3), see Figure 4. On the other hand, including contact duration (Model 3) does not have a clear-cut effect as compared to the model considering the number of contacts by age only (Model 2).

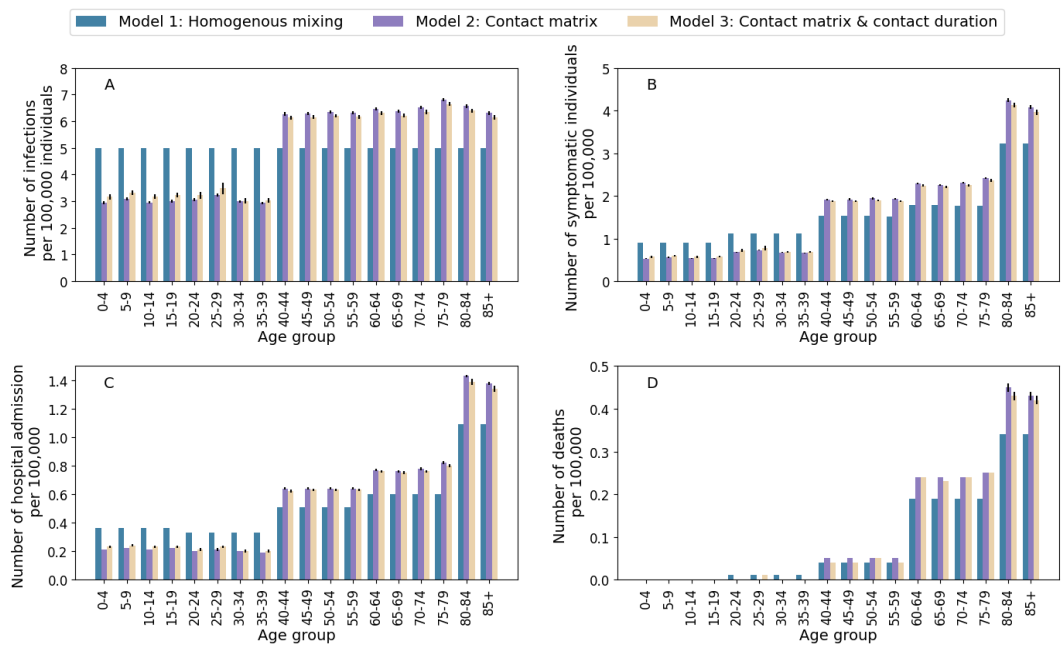

**Figure 4**: **Application to COVID-19. A** Number of infections by age group for the three models. $R_0$ is set to 1.3, the simulation is interrupted when 1,000 infections are reached in a population of 20 million individuals. **B** As A, but for symptomatic individuals. **C** As A, but for hospital admissions. **D** As A, but for deaths.

## 4. Discussion

In this paper, we analyzed contact intensity from two different angles: contact number and average contact duration. Overall, individuals reported an average of 2.3 contacts per day and 7.0 hours per contact. The estimated contact pattern by age is consistent with the other estimates obtained for China during the COVID-19 pandemic [7,15].

In the post-lockdown period, when workers had started to resume in-person work, non-negligible percentages of contact number and contact duration were recorded in the workplace with colleagues. However, contacts reported with family members at home still represented a remarkable fraction of overall recorded contacts and contact duration, indicating that during the study period people were still cautious about resuming contacts.

Our modeling exercise highlight the effect of considering human mixing patterns when estimating COVID-19 burden. Considering contact duration appear to have a limited effect on the estimation of COVID-19 burden, but further investigations are warranted.

The following limitations need to be considered when interpreting the results presented in this study. The current survey over-samples school-age individuals and participants needed to be able to have access to the internet, potentially biasing our results. Recorded contacts can be affected by recall bias and compliance bias as well. Most schools were still closed (online learning) during the study period; this does not allow us to draw conclusions about the post-lockdown mixing patterns of Chinese students. This study is based on limited number of participants (354) that, although larger than the sample used for the Netherlands in the POLYMOD study [10], does not allow us to consider refined age groups. In particular, we did not have enough sample to subdivide individuals aged 40+ years in multiple subgroups. As such, we are not able to draw specific conclusions for the contact patterns followed by the elderly – a key group in determining COVID-19 burden. Finally, the modeling exercise presented here does not have the ambition of being representative of the all the complexities of COVID-19 epidemiology; it serves as a simplified example to highlight the effect of mixing patterns in SARS-CoV-2 transmission patterns and COVID-19 burden.

In light of these limitations, this work can be considered as an incremental step in quantifying contact patterns by age, relationship, social setting, and duration in China during the COVID-19 pandemic.

**Funding**

This research was supported by the National Natural Science Foundation of China under Grant No. 62003230, the Fundamental Research Funds for the Central Universities under Grant No. 1082204112289, the Special Funds for Prevention and Control of COVID-19 of Sichuan University under Grant No. 0082604151026.

**Declaration of competing interest**

M.A. has received research funding from Seqirus. The funding is not related to COVID-19. All other authors declare no competing interest.

**References**
[1] Chinese Center for Disease Control and Prevention. Notice on the implementation of science-based and targeted measures in the prevention and control of novel coronavirus pneumonia. 2020. Available in Chinese at: http://www.chinacdc.cn/jkzt/crb/zl/szkb_11803/jszl_11815/202002/t20200225_213723.html (Accessed on Nov. 25, 2021).
[2] Zhang J, Litvinova M, Wang W, et al. Evolving epidemiology and transmission dynamics of coronavirus disease 2019 outside Hubei province, China: a descriptive and modelling study. The Lancet Infectious Diseases, 2020, 20(7): 793-802.
[3] Jia J, Lu X, Yuan Y, et al. Population flow drives spatio-temporal distribution of COVID-19 in China. Nature, 2020, 582(7812): 389-394.
[4] Latsuzbaia A, Herold M, Bertemes J P, et al. Evolving social contact patterns during the COVID-19 crisis in Luxembourg. PLOS one, 2020, 15(8): e0237128.
[5] Feehan D M, Mahmud A S. Quantifying population contact patterns in the United States during the COVID-19 pandemic. Nature Communications, 2021, 12(1): 1-9.
[6] Backer J A, Mollema L, Vos E R A, et al. Impact of physical distancing measures against COVID-19 on contacts and mixing patterns: repeated cross-sectional surveys, the Netherlands, 2016–17, April 2020 and June 2020. Eurosurveillance, 2021, 26(8): 2000994.
[7] Zhang J, Litvinova M, Liang Y, et al. Changes in contact patterns shape the dynamics of the covid-19 outbreak in china. Science, 2020, 368(6498):1481-1486.
[8] Jarvis C I, Van Zandvoort K, Gimma A, et al. Quantifying the impact of physical distance measures on the transmission of COVID-19 in the UK. BMC Medicine, 2020, 18(1): 1-10.
[9] Liu C Y, Berlin J, Kiti M C, et al. Rapid review of social contact patterns during the COVID-19 pandemic. Epidemiology, 2021, 32(6): 781-791.
[10] Mossong J, Hens N, Jit M, et al. Social contacts and mixing patterns relevant to the spread of infectious diseases. PLOS Medicine, 2008, 5(3): e74.
[11] Kretzschmar M, Mikolajczyk R T. Contact profiles in eight European countries and implications for modelling the spread of airborne infectious diseases. PLOS One, 2009, 4(6): e5931.
[12] Leung K, Jit M, Lau E H Y, et al. Social contact patterns relevant to the spread of respiratory infectious diseases in Hong Kong. Scientific Reports, 2017, 7(1): 1-12.
[13] Ajelli M, Litvinova M. Estimating contact patterns relevant to the spread of infectious diseases in Russia. Journal of Theoretical Biology, 2017, 419:1-7.
[14] Zhang J, Klepac P, Read J M, et al. Patterns of human social contact and contact with animals in Shanghai, China. Scientific reports, 2019, 9(1): 1-11.
[15] Zhang J, Litvinova M, Liang Y, et al. The impact of relaxing interventions on human contact patterns and SARS-CoV-2 transmission in China. Science Advances, 2021, 7(19):eabe2584.

[16] Kraemer M U G, Yang C H, Gutierrez B, et al. The effect of human mobility and control measures on the COVID-19 epidemic in China. Science, 2020, 368(6490): 493-497.

[17] Schlosser F, Maier B F, Jack O, et al. COVID-19 lockdown induces disease-mitigating structural changes in mobility networks. Proceedings of the National Academy of Sciences, 2020, 117(52): 32883-32890.

[18] Kogan N E, Clemente L, Liautaud P, et al. An early warning approach to monitor COVID-19 activity with multiple digital traces in near real time. Science Advances, 2021, 7(10): eabd6989.

[19] Questionnaire Star. Company webpage in Chinese: https://www.wjx.cn/. (Accessed on Nov. 25, 2021).

[20] National Bureau of Statistics of the People's Republic of China. China Statistical Yearbook 2010. Available in Chinese at http://www.stats.gov.cn/tjsj/ndsj/2010/indexch.htm. (Accessed on Nov. 25, 2021).

[21] Efron B. Bootstrap methods: another look at the jackknife. Breakthroughs in Statistics. Springer, New York, NY, 1992: 569-593.

[22] Anderson R M, May R M. Infectious Diseases of Humans: dynamics and Control, Oxford University Press, Oxford, UK, 1992.

[23] Wallinga J, Lipsitch M. How generation intervals shape the relationship between growth rates and reproductive numbers. Proceedings of the Royal Society B: Biological Sciences, 2007, 274(1609): 599-604.

[24] Liu Q H, Ajelli M, Aleta A, et al. Measurability of the epidemic reproduction number in data-driven contact networks. Proceedings of the National Academy of Sciences, 2018, 115(50): 12680-12685.

[25] Diekmann O, Heesterbeek J A P, Metz J A J. On the definition and the computation of the basic reproduction ratio $R_0$ in models for infectious diseases in heterogeneous populations. Journal of Mathematical Biology, 1990, 28(4): 365-382.

[26] Poletti P, Tirani M, Cereda D, et al. Association of age with likelihood of developing symptoms and critical disease among close contacts exposed to patients with confirmed sars-cov-2 infection in italy. JAMA Network Open, 2021, 4(3): e211085-e211085.

[27] Yang J, Chen X, Deng X, et al. Disease burden and clinical severity of the first pandemic wave of COVID-19 in Wuhan, China. Nature Communications, 2020, 11(1): 1-10.

[28] Trentini F, Guzzetta G, Galli M, et al. Modeling the interplay between demography, social contact patterns, and SARS-CoV-2 transmission in the South West Shewa Zone of Oromia Region, Ethiopia. BMC Medicine, 2021, 19(1): 1-13.

[29] Béraud G, Kazmerczak S, Beutels P, et al. The French connection: the first large population-based contact survey in France relevant for the spread of infectious diseases. PLOS One, 2015, 10(7): e0133203.

**Appendix A**

**Table S1. Parameters regulating COVID-19 burden.**

| Description | Age (years) | Value (%) | Reference |
|---|---|---|---|
| Probability of developing respiratory symptoms and/or fever | 0-19 | 18.1 | Poletti et al. [26] |
| | 20-39 | 22.4 | |
| | 40-59 | 30.5 | |
| | 60-79 | 35.5 | |
| | 80+ | 64.6 | |
| Proportion of laboratory-confirmed symptomatic individuals requiring hospitalization by age *a* | 0-19 | 40.0 | Yang et al. [27] |
| | 20-39 | 29.2 | |
| | 40-59 | 33.3 | |
| | 60+ | 33.8 | |
| Fatality ratio among laboratory-confirmed symptomatic individuals | 0-19 | 0.51 | Yang et al. [27] |
| | 20-39 | 0.65 | |
| | 40-59 | 2.38 | |
| | 60+ | 10.52 | |

**Table S2. Data completeness.**

| Province | Total Number of Participants | Number of Participants with Missing Information | Number of Participants with Complete Information |
|---|---|---|---|
| **Sichuan** | 360 | 160 | 200 |
| **Chongqing** | 34 | 14 | 20 |
| **Shandong** | 33 | 14 | 19 |
| **Hebei** | 23 | 6 | 17 |
| **Henan** | 25 | 9 | 16 |
| **Zhejiang** | 21 | 7 | 14 |
| **Fujian** | 18 | 5 | 13 |
| **Yunnan** | 23 | 10 | 13 |
| **Hunan** | 19 | 7 | 12 |
| **Guangdong** | 19 | 9 | 10 |
| **Jiangxi** | 17 | 7 | 10 |
| **Shanxi** | 21 | 11 | 10 |
| **Total** | 613 | 259 | 354 |

One participant was not a Chinese resident and one responded to the survey outside of the 20-day study period; as such, they were excluded from the study.